\documentclass[superscriptaddress]{iopart}

\usepackage{graphicx}
\usepackage{dcolumn}
\usepackage{bm}

\newcommand{\bo}{\raise-1mm\hbox{\Large$\Box$}}

\newcommand{\rrho}{\boldsymbol{\rho}}
\newcommand{\llambda}{\boldsymbol{\lambda}}
\newcommand{\mmu}{\boldsymbol{\mu}}
\newcommand{\xx}{\mathbf{x}}

\newcommand{\qq}{\mathbf{q}}

\newcommand{\dd}{\mathbf{d}}
\newcommand{\EE}{\mathbf{E}}
\usepackage{epsfig}
\usepackage{latexsym}
\usepackage{mathrsfs}
\usepackage{setspace}
\usepackage{amssymb}
\usepackage{wrapfig}
\usepackage{color}

\begin{document}

\title{A dielectric superfluid of polar molecules}

\author{Ryan M. Wilson$^1$, Seth T.\ Rittenhouse$^2$ and John L. Bohn$^1$}
\address{$^1$JILA and Department of Physics, University of Colorado, Boulder, Colorado 80309-0440, USA}
\address{$^2$ITAMP, Harvard-Smithsonian Center for Astrophysics, Cambridge, MA 02138, USA}
\ead{rmw@colorado.edu}

\begin{abstract}
We show that, under achievable experimental conditions, a Bose-Einstein condensate (BEC) of polar molecules can exhibit dielectric character.  In particular, we derive a set of self-consistent mean-field equations that couple the condensate density to its electric dipole field, leading to the emergence of polarization modes that are coupled to the rich quasiparticle spectrum of the condensate.  While the usual roton instability is suppressed in this system, the coupling can give rise to a phonon-like instability that is characteristic of a dielectric material with a negative static dielectric function.  
\end{abstract}
\date{\today}
\maketitle


The electrical properties of any material are encoded in its dielectric function $\kappa ({\bf q},\omega)$, which relates the induced dipole moments within the material to an applied electric field with wave vector $\qq$ and frequency $\omega$.  For its predicted relationship to a variety of novel physical phenomena, the possibility of $\kappa$ taking on negative values has generated much interest.  For the ac case, it is well known that the dielectric function can become negative, for example, at frequencies near an atomic resonance, resulting in the high reflectivity of metals~\cite{JacksonEM}.  Similarly, under specially engineered circumstances the magnetic permeability of certain materials can become negative, resulting in a negative index of refraction~\cite{Veselago68,Shelby01}.

For a dc field, however, the existence of a negative dielectric function becomes more subtle.  Formally, in the infinite-wavelength limit ${\bf q} \rightarrow 0$, a material is stabilized only if its dielectric function, or constant, is non-negative~\cite{Dolgov81}.  Such is not necessarily the case for ${\bf q} \neq 0$, however, where the dielectric constant may take on negative values without violating causality~\cite{Iwamoto91}.  The presence of a negative dielectric constant ($\kappa(\qq,\omega=0)<0$) is predicted to significantly alter the electrical properties of materials by, for example, giving rise to the attraction of like charges within a material.  In this case, the negative dielectric constant is thought to have profound implications for the maximum critical temperature for superconductivity~\cite{Dolgov81}.  Experimentally testing such assertions has proven difficult, although the dc limit has been approached in certain materials~\cite{Shulman09}.

In this article, we propose that the realm of negative dielectric constant can be studied in soon-to-be constructed ultracold molecule experiments, in which electrically polar molecules are produced in a Bose-Einstein condensate (BEC).  Specifically, we consider molecules whose permanent dipole moments are easily aligned in sufficiently weak applied fields.  In this case, the molecules exert fields on one another that are comparable to the applied field, and the BEC acquires dielectric as well as superfluid character.  
We show that altering either the density of the gas or the strength of the applied field can drive the dielectric constant of the system to be negative, triggering a long-wavelength instability of the condensate.   This instability is qualitatively completely different than those that have been studied previously in dipolar BECs with fixed dipole moments, being the phonon~\cite{Nath09} and roton instabilities~\cite{Santos03}.  Additionally, the decay times of this dielectric instability are found to be relatively long, allowing for this state to be observed and studied in detail.  Critical in our analysis is the emergence of polarization modes that are coupled to quasiparticle excitations of the BEC.  Thus, we show that nontrivial physics emerges in the system of ultracold polar molecules in small fields, and motivate the exploration of this regime, or the regime of the dielectric superfluid, for future experiments.

As a qualitative guide, we first consider a homogeneous dielectric placed in an applied field.  This system is characterized by the well-known Clausius-Mossotti relation, which expresses the dielectric constant $\kappa$ in terms of the polarizability $\alpha$ of the microscopic constituents,
\begin{equation}
\label{CM}
\kappa-1 = \frac{n\alpha}{1-\frac{4\pi}{3}n\alpha}
\end{equation}
where $n$ is the number density of the dielectric~\cite{JacksonEM}.  Eq.~(\ref{CM}) suggests that large enough values of $n\alpha$ could drive the dielectric constant of the homogeneous material to be negative.  As discussed in~\cite{Dolgov81}, this is indeed possible for materials with static dielectric functions $\kappa(\qq,\omega=0)$ with small but non-zero wave-number ($q>0$).  Most dielectric materials such as ceramics, glasses and plastics have $\kappa \sim 1-10$, mostly due to their large densities~\cite{CRC}, while most gases have $\kappa \sim 1$ because their densities are very small.  Gases of heteronuclear polar molecules, however, can realize large values of $n\alpha$ due to their possibly large polarizabilities, resulting in very large dielectric constants for these systems even in the dilute gaseous state where $n$ is small.

The heteronuclear molecules that we consider here possess permanent dipolar moments $\mathrm{d}_\mathrm{max}$ in their body-fixed frames.  In the absence of an applied electric field, the ground state of each molecule is a parity eigenstate and the orientations of the molecular dipole moments average to zero in the laboratory frame.  However, the presence of an applied electric field mixes the molecular parity states, and the average dipole orientation becomes nonzero.  Because the energy splitting between these states, $\Delta$, can be very small, the resulting molecular polarizabilities can be very large (see table~\ref{Candidatemolecules} for examples).  In particular, this polarizability is distinct from the more familiar polarizability, which is on the order of 1-100 a.u., and arises solely from the electric degrees of freedom.  In a weak electric field $\EE$, the molecules then acquire a laboratory-frame dipole moment $\dd = \alpha \EE$ characterized by the molecular polarizability $\alpha \propto d_\mathrm{max}^2 / \Delta$, where $d_\mathrm{max}$ is the limiting value of the molecular dipole moment in large applied fields and $\Delta$ is the splitting of the parity doublet~\cite{FriedrichTAP}.  This polarizability is an intrinsic property of a heteronuclear molecule.  The molecular dipole moment in the laboratory frame is therefore uniquely determined by the local electric field via the polarizability $\alpha$.

While the gas of heteronuclear polar molecules can possess a large polarizability due to the small parity splittings, this feature is quickly lost at finite temperature.  Indeed, a \emph{thermal} gas of polar molecules, at temperatures $T$ such that $k_B T > \Delta$, has both parity states equally populated, causing the dipolar character of the system to vanish.  However, in an ultracold, Bose-condensed gas the lower molecular state can be dominantly populated, thus realizing high polarizability $\alpha$.  Thus, polar molecules, at sufficiently low temperature, can exhibit polarizability and hence dielectric character, albeit from a different microscopic origin than atoms or non-polar molecules.  

To deal with the properties of this inhomogeneous dielectric system, we cannot simply employ the results derived for a homogeneous material.  We therefore begin by considering a generalized dielectric formalism.  The energy density of a dielectric material with a polarization $\mathbf{P}(\xx)$ induced by an electric field $\mathbf{E}(\xx)$ is given by
\begin{equation}
\label{w}
w(\xx) = -\frac{1}{2}\mathbf{P}(\xx)\cdot \mathbf{E}(\xx).
\end{equation}
To account fully for the internal interaction energy of the system, we write the electric field as $\mathbf{E}(\xx) = \mathbf{E}_\mathrm{app} + \mathbf{E}_\mathrm{P}(\xx)$, where $\EE_\mathrm{app}$ is the applied, or external field in the absence of the BEC and $\mathbf{E}_\mathrm{P}(\xx)$ is the polarization field.  The polarization, by definition, is the dipole density $\mathbf{P}(\xx) = n(\xx)\mathbf{d}(\xx)$, where $n(\xx)$ is the particle density of the dielectric and $\mathbf{d}(\xx)$ is the local dipole moment of a constituent at point $\xx$, or the dipole field.

In a dielectric material, the dipoles are induced by the local field and can be expressed, in the linearly polarizable regime, as $\mathbf{d}(\xx) = \alpha \mathbf{E}(\xx)$.  This leads to an expression for the polarization of the dielectric,
\begin{equation}
\label{P}
\mathbf{P}(\xx) = n(\xx) \alpha \left[ \mathbf{E}_\mathrm{app} + \mathbf{E}_\mathrm{P}(\xx) \right].
\end{equation}
We can now calculate the field produced at a point $\xx$ due to the presence of this polarization simply by the convolution over the electric field produced by a dipole at a point $\xx^\prime$,
\begin{equation}
\label{EP}
\mathbf{E}_\mathrm{P}(\xx) = \int d\xx^\prime \frac{3 \,\mathbf{n} \left( \mathbf{P}(\xx^\prime)\cdot\mathbf{n}\right) - \mathbf{P}(\xx^\prime)}{|\xx-\xx^\prime|^3},
\end{equation}
where $\mathbf{n}$ is the unit vector in the direction $\xx-\xx^\prime$.  A self consistent solution to Eqs.~(\ref{P}) and~(\ref{EP}) can be used to calculate the energy density of a dielectric material, Eq.~(\ref{w}), given the density $n(\xx)$ and polarizability $\alpha$ of the constituents.

For a Bose-condensed gas of polar molecules in the dilute regime, we interpret the dielectric density as the magnitude squared of the mean-field condensate wave function, $n(\xx) = |\Psi(\xx)|^2$, where $\int d\xx |\Psi(\xx)|^2 = N$, and $N$ is the number of molecules in the system.  Now, the polarization can be expressed as $\mathbf{P}(\xx) = \dd(\xx) |\Psi(\xx)|^2$ and the field $\EE_\mathrm{P}(\xx)$ in Eq.~(\ref{EP}) can be expressed in terms of the dipole field $\dd(\xx)$ and the wave function $\Psi(\xx)$,
\begin{equation}
\label{EP2}
\EE_\mathrm{P}(\xx) = - \int d\xx^\prime \frac{ \dd(\xx^\prime) - 3\,\mathbf{n}\left( \dd(\xx^\prime) \cdot \mathbf{n} \right)}{|\xx-\xx^\prime|^3}  |\Psi(\xx^\prime)|^2.
\end{equation}
The dipole field $\dd(\xx)$ gives the dipole moment of a molecule at point $\xx$.  From Eqs.~(\ref{P}) and~(\ref{EP2}), it follows that the dipole field obeys the self-consistent expression depending on $\Psi(\xx)$,
\begin{equation}
\label{dfieldgen}
\dd(\xx) = \alpha\left[ \EE_\mathrm{app} - \int d\xx^\prime \frac{ \dd(\xx^\prime) - 3\,\mathbf{n}\left( \dd(\xx^\prime) \cdot \mathbf{n} \right)}{|\xx-\xx^\prime|^3}  |\Psi(\xx^\prime)|^2 \right],
\end{equation}
where we see that a molecule with polarizability $\alpha$ at point $\xx$ obtains an induced dipole moment $\dd(\xx)$ due to contributions from both the applied field $\EE_\mathrm{app}$ and the field produced by the other dipoles in the system.

From Eq.~(\ref{w}), we can now express the energy of this system, including kinetic and trapping energies, as a functional of $\Psi(\xx)$ and the dipole field $\dd(\xx)$.  Neglecting any short-range, non-dipolar interactions between the molecules, this energy functional is given by
\begin{eqnarray}
\label{bigE}
E[\Psi,\dd] &=& \int d\xx \, \Psi^\star(\xx) \left[  -\frac{\hbar^2}{2M} \nabla^2 + U(\xx) - \frac{1}{2} \dd(\xx)\cdot \bigg[ \EE_\mathrm{app} \right. \nonumber \\
 &-&  \left. \int d\xx^\prime \frac{ \dd(\xx^\prime) - 3\,\mathbf{n}\left( \dd(\xx^\prime) \cdot \mathbf{n} \right)}{|\xx-\xx^\prime|^3}  |\Psi(\xx^\prime)|^2 \bigg] \right]  \Psi(\xx),
\end{eqnarray}
where $M$ is the mass of a polar molecule and $U(\xx)$ is the trapping potential.  The ground state of this system is found by minimizing the total energy with respect to any variation of $\Psi^\star$.    Such a task is seemingly straightforward, however, care must be taken when considering the $\Psi(\xx)$ dependence of the dipole field $\dd(\xx)$.  Due to the self-consistent nature of Eq.~(\ref{dfieldgen}), a functional derivative of $\dd(\xx)$ with respect to $\Psi^\star(\xx)$ is not straightforward.  To handle this relationship, we treat Eq.~(\ref{dfieldgen}) as a constraint on the system and employ a Lagrange multiplier technique for minimizing~(\ref{bigE}).

We define a modified energy functional $\Omega[\Psi,\dd]$, given by
\begin{equation}
\label{biglambda}
\Omega[\Psi,\dd] = E[\Psi,\dd] + \llambda \cdot \mathbf{F} [\Psi,\dd], 
\end{equation}
where $\mathbf{F}[\Psi,\dd]$ is defined so that Eq.~(\ref{dfieldgen}) is satisfied when $\partial \Omega / \partial \lambda_i = 0$ and, by construction, $\llambda$ is a vector Lagrange multiplier with units of electric field.  Now, a wave function $\Psi(\xx)$ that minimizes the total energy of this system is found by setting $\partial \Omega / \partial \Psi^\star = 0$.  Performing this operation, we derive the non-linear Schr\"odinger Equation, or Gross-Pitaevskii equation (GPE) for $\Psi(\xx)$, given by
\begin{eqnarray}
\label{gpe1}
\mu \Psi(\xx) &=& \left[ -\frac{\hbar^2}{2M} \nabla^2 + U(\xx) + \frac{1}{2}\dd(\xx)\cdot  \left( \llambda - \EE_\mathrm{app}  \right)   \right. \nonumber \\
&+& \! \! \left. \left( \alpha \llambda + \dd(\xx) \right) \cdot \! \! \int d\xx^\prime \frac{ \dd(\xx^\prime) - 3\,\mathbf{n}\left( \dd(\xx^\prime) \cdot \mathbf{n} \right)}{|\xx-\xx^\prime|^3}  |\Psi(\xx^\prime)|^2   \right] \Psi(\xx),
\end{eqnarray}
where $\mu = \mu_E + \llambda \cdot \mmu_F$ is the chemical potential of the condensate, whose terms $\mu_E$ and $\mu_{F,i}$ can be calculated by projecting Eq.~(\ref{gpe1}) onto $\Psi^\star(\xx)$ and matching the terms proportional to $1$ and $\lambda_i$, respectively.  The components  $\lambda_i$ of the vector Lagrange multiplier are found by enforcing $\partial \Omega / \partial \mathrm{d}_i = 0$, and obey the coupled equations
\begin{eqnarray}
\label{smalllambda}
&{}& \frac{1}{2} N \mathrm{E}_{\mathrm{app},i} - \int d\xx |\Psi(\xx)|^2 \int d\xx^\prime |\Psi(\xx^\prime)|^2 \mathrm{d}_i(\xx^\prime) \frac{1-3\cos^2{\theta_i}}{|\xx-\xx^\prime|^3} \nonumber \\
&{}& +  \sum_{\gamma = j,k} \int d\xx |\Psi(\xx)|^2 \int d\xx^\prime |\Psi(\xx^\prime)|^2 \mathrm{d}_\gamma(\xx^\prime) \frac{3\cos{\theta_i} \cos{\theta_\gamma}}{|\xx-\xx^\prime|^3} \nonumber \\
&{}& =  \frac{1}{2}\lambda_i \left[ N+\alpha \int d\xx |\Psi(\xx)|^2 \int d\xx^\prime |\Psi(\xx^\prime)|^2 \frac{1-3\cos^2{\theta_i}}{|\xx-\xx^\prime|^3}  \right] \nonumber \\
&{}& - \frac{1}{2} \sum_{\gamma = j,k} \lambda_\gamma  \alpha \int d\xx |\Psi(\xx)|^2 \int d\xx^\prime |\Psi(\xx^\prime)|^2 \frac{3\cos{\theta_i} \cos{\theta_\gamma}}{|\xx-\xx^\prime|^3},
\end{eqnarray}
where $\theta_i$ is the angle between $\mathbf{n}$ and $\hat{i}$.  Equations for the components $\lambda_j$ and $\lambda_k$ are found by cyclic permutations of the indices in Eq.~(\ref{smalllambda}).  Thus, the ground state of the dilute BEC of polar molecules in weak applied fields is described by Eqs.~(\ref{dfieldgen}),~(\ref{gpe1}) and~(\ref{smalllambda}).  Interestingly, we see that these equations couple the condensate wave function to the dipole field, so a change in density can result in a change in the dipole moment of the polar molecules, and vice versa.  In the strong field limit of fully polarized dipoles, however, $\dd(\xx)\rightarrow \dd_\mathrm{max}$ and $\alpha \rightarrow 0$.  In this case, Eq.~(\ref{gpe1}) becomes identical to the non-local GPE that is typically used to study dipolar BECs~\cite{Lahaye09Rev}.

An instructive and relevant case to consider is a gas of polar molecules in a one-dimensional (1D) harmonic trap, where $U(\xx) = \frac{1}{2}M\omega_z^2 z^2$ and the applied field is homogeneous in the $z$-direction, $\EE_\mathrm{app} = \hat{z} \mathrm{E}_\mathrm{app}$.  Indeed, such quasi-two dimensional (q2D) geometries are sought after to stabilize dipolar gases against collisional loss~\cite{Buchler07,TicknorRittenhouse10,NiNature10} and energetic instabilities due to the attractive part of the dipole-dipole interaction~\cite{Koch08}.  In this translationally invariant geometry, all transverse (in-plane) components of the electric field due to the dipole polarization will cancel, so we can write $\EE_\mathrm{P}(\xx) = \hat{z}\mathrm{E}_\mathrm{P}(\xx)$, and subsequently $\dd(\xx) = \hat{z} \mathrm{d}(\xx)$.  With these simplifications, the GPE (Eq.~(\ref{gpe1})) can be rewritten as
\begin{eqnarray}
\label{gpe2}
\mu \Psi(\xx) &=& \left[ -\frac{\hbar^2}{2M} \nabla^2 + U(\xx) + \frac{1}{2} \mathrm{d}(\xx) \left(\lambda -\mathrm{E}_\mathrm{app} \right) \right. \nonumber \\
&+& \left. \left( \alpha \lambda + \mathrm{d}(\xx) \right)  \int d\xx^\prime  \mathrm{d}(\xx^\prime) |\Psi(\xx^\prime)|^2    \frac{ 1 - 3\cos^2{\theta_z}}{|\xx-\xx^\prime|^3} \right] \Psi(\xx),
\end{eqnarray}
and the equation for the dipole field (Eq.~(\ref{dfieldgen})) can be rewritten as
\begin{equation}
\label{Psym}
\mathrm{d}(\xx) = \alpha \left[ \mathrm{E}_\mathrm{app} - \int d\xx^\prime |\Psi(\xx^\prime)|^2 \mathrm{d}(\xx^\prime) \frac{1-3\cos^2{\theta_z}}{|\xx-\xx^\prime|^3} \right],
\end{equation}
where $\theta_z$ is the angle between $\mathbf{n}$ and $\hat{z}$.  In this geometry, the vector Lagrange multiplier has just one non-zero component corresponding to the $z$-direction, $\lambda = \hat{z} \cdot \llambda$, where
\begin{equation}
\label{lambda}
\lambda = \frac{N \mathrm{E}_\mathrm{app} - 2\int d\xx |\Psi(\xx)|^2 \int d\xx^\prime |\Psi(\xx^\prime)|^2 \mathrm{d}(\xx^\prime) \frac{1-3\cos^2{\theta_z}}{|\xx-\xx^\prime|^3} }{N+\alpha \int d\xx |\Psi(\xx)|^2 \int d\xx^\prime |\Psi(\xx^\prime)|^2 \frac{1-3\cos^2{\theta_z}}{|\xx-\xx^\prime|^3}}.
\end{equation}
Eqs.~(\ref{gpe2}),~(\ref{Psym}) and~(\ref{lambda}) fully describe the BEC of polar molecules in the q2D geometry described above.  Thus, we seek solutions for the condensate wave function $\Psi(\xx)$ and the corresponding dipole field $\mathrm{d}(\xx)$ that self-consistently satisfy these equations.  We note that the results presented in the following are nearly quantitatively equivalent to those that follow from neglecting the $\Psi(\xx)$ dependence of $\dd(\xx)$ in the derivation of the GPE, Eq.~(\ref{gpe1}), which amounts to setting $\lambda=0$ in this work.  The governing equations for this system are, however, quite general for describing a BEC of polar molecules in weak applied fields, and may be crucial for future studies.

To describe the \emph{pure} condensate in this q2D geometry, we note that the lowest energy state is that with zero in-plane momentum, $\Psi(\xx)\rightarrow \Psi(z) = \sqrt{n_\mathrm{2D}} \chi(z)$, where $n_\mathrm{2D}$ is an integrated 2D density and $\chi(z)$ is the axial wave function of the condensate, normalized to unity.  For a condensate wave function with this form, the polarization field, and therefore the dipole field of the condensate, depend only on $z$.  For this case, all of the convolution integrals in Eqs.~(\ref{gpe2}),~(\ref{Psym}) and~(\ref{lambda}) can be performed analytically, using the result from Ref.~\cite{Santos03}
\begin{equation}
\label{convexamp}
\int d\xx^\prime f(z^\prime) \frac{1-3\cos^2{\theta_z}}{|\xx-\xx^\prime|^3} = \frac{8\pi}{3} f(z).
\end{equation}
Thus, we see that Eq.~(\ref{Psym}) reduces to
\begin{equation}
\label{d0z2}
\mathrm{d}(z) = \alpha \left[  \mathrm{E}_\mathrm{app} - \frac{8\pi n_\mathrm{2D}}{3} |\chi(z)|^2 \mathrm{d}(z)\right],
\end{equation}
which yields the solution for the condensate dipole field,
\begin{equation}
\label{dz}
\mathrm{d}(z) = \frac{\alpha \mathrm{E}_\mathrm{app}}{1+\frac{8\pi n_\mathrm{2D} \alpha}{3} |\chi(z)|^2}.
\end{equation}
The form of this result is in stark contrast with the well-known Clausius-Mossotti result~(\ref{CM}), where the denominator of Eq.~(\ref{dz}) is replaced by $1-\frac{4\pi}{3} n\alpha$, where $n$ is the density of the homogeneous 3D dielectric.  Indeed, the Clausius-Mossotti result does not necessarily hold for inhomogeneous systems, like the one we consider here, which has a finite extent in the polarization direction.  For this case, the net electric fields of the other dipolar molecules in the system sum to point locally in the direction opposing the applied field.  For linearly polarizable molecules, this results in a decrease of the local dipole moment, which is reflected in Eq.~(\ref{dz}).  The case is the opposite for the homogeneous 3D system, or a quasi-2D system where the applied field lies parallel to the plane of symmetry, or indeed any system that is homogeneous in the direction of the applied or polarization field.  In this case, the result~\cite{Santos03}
\begin{equation}
\label{convexampxy}
\int d\xx^\prime f(x^\prime,y^\prime) \frac{1-3\cos^2{\theta_z}}{|\xx-\xx^\prime|^3} = -\frac{4\pi}{3} f(x,y)
\end{equation}
can be used to reproduce the Clausius-Mossotti result.

\begin{figure}
\centering
\includegraphics[width=0.8\textwidth]{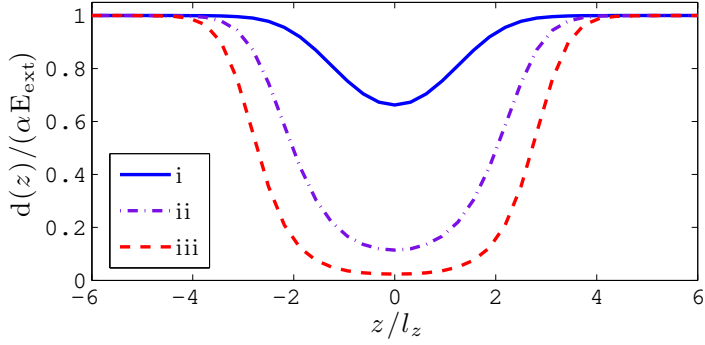} \\
\caption{(color online)  The dipole field corresponding to the condensate mode of the q2D system, given by Eq.~(\ref{dz}).  For this figure, we take the axial wave function $\chi(z)$ to be a Gaussian normalized to unity with width $l_z$.   The lines correspond to i) $n_\mathrm{2D} \alpha |\chi(0)|^2=.046$, ii) $n_\mathrm{2D}\alpha |\chi(0)|^2=0.70$ and iii) $n_\mathrm{2D}\alpha |\chi(0)|^2=3.71$, and coincide with the points shown in figure~\ref{fig:stab}.  For a given polarizability, higher densities reduce the local dipole moment inside the gas.}
\label{fig:dfield}
\end{figure}

For the case at hand, however, Eq.~(\ref{dz}) makes clear the interesting relation between the condensate density and the corresponding dipole field, revealing that the dipole moments of the condensed molecules are smaller where their density is larger.  This dipole field is shown in figure~\ref{fig:dfield} for various values of $n_\mathrm{2D} \alpha |\chi(0)|^2$, where we take $\chi(z)$ to be a Gaussian of width $l_z = \sqrt{\hbar / M \omega_z}$.  The widely studied dipolar BECs with fixed dipole moments, trapped in oblate or q2D geometries, are predicted to be unstable to roton-like quasiparticles beyond some critical density or interaction strength set by the dipole moment of the bosons~\cite{Santos03,Ronen07}.  Eq.~(\ref{dz}) suggests that this mechanism may be suppressed in the dielectric BEC for larger values of $n \alpha $ as the dipole moments are diminished in the region of higher density.


We explore the stability of this system more quantitatively by investigating the behavior of small fluctuations on top of the ground state, that is, when $\Psi(z) \rightarrow \Psi(z) + \delta \psi(\xx,t)$, and $\delta \ll 1$.  Such perturbations to the ground condensed state do not necessarily possess translational symmetry, and a full treatment of the mean-field Eqs.~(\ref{gpe2}) and~(\ref{Psym}) must be considered.  To handle the time-dependence of the fluctuations $\psi(\xx,t)$, we generalize Eq.~(\ref{gpe2}) to its time-dependent form, where the stationary condensate has the time dependence $e^{-i\mu t}$.  In this case, $\mu$ is replaced by $i \hbar \partial_t$ in Eq.~(\ref{gpe2}).

Whereas the usual linearization of the GPE results in a single equation for $\psi(\xx,t)$, the linearization of Eqs.~(\ref{gpe2}) and~(\ref{Psym}) results in an equation for $\psi(\xx,t)$ and an equation for the corresponding linear response of the dipole field.  To this end, we take $\mathrm{d}(z) \rightarrow \mathrm{d}(z) + \delta \sigma(\xx,t)$, where $\sigma(\xx,t)$ describes the linear deviations from the condensate dipole field $d(z)$ due to the fluctuations $\psi(\xx,t)$.  The equation for $\psi(\xx,t)$ is given by
\begin{eqnarray}
\label{dpsieq}
i\hbar \partial_t \psi(\xx,t) &=& \left[ -\frac{\hbar^2}{2M}\nabla^2 + U(\xx) - \mu  \right]\psi(\xx,t) \nonumber \\
&+& \left( \alpha \lambda + \mathrm{d}(z) \right) \left[ \int d\xx^\prime \frac{1-3\cos^2{\theta_z}}{|\xx-\xx^\prime|^3} \sigma(\xx^\prime,t) |\Psi(z^\prime)|^2  \right] \Psi(z) \nonumber \\
&+& \left( \alpha \lambda + \mathrm{d}(z) \right) \left[ \int d\xx^\prime \frac{1-3\cos^2{\theta_z}}{|\xx-\xx^\prime|^3} \mathrm{d}(z^\prime) \Psi^\star(z^\prime) \psi(\xx^\prime,t)  \right] \Psi(z) \nonumber \\
&+& \left(  \alpha \lambda +\mathrm{d}(z) \right) \left[ \int d\xx^\prime \frac{1-3\cos^2{\theta_z}}{|\xx-\xx^\prime|^3} \mathrm{d}(z^\prime) \Psi(z^\prime) \psi^\star(\xx^\prime,t)  \right] \Psi(z) \nonumber \\
&+& \left( \alpha \lambda + \mathrm{d}(z) \right) \left[ \int d\xx^\prime \frac{1-3\cos^2{\theta_z}}{|\xx-\xx^\prime|^3} \mathrm{d}(z^\prime) |\Psi(z^\prime)|^2  \right] \psi(\xx,t) \nonumber \\
&+& \sigma(\xx,t) \left[ \int d\xx^\prime \frac{1-3\cos^2{\theta_z}}{|\xx-\xx^\prime|^3} \mathrm{d}(z^\prime) |\Psi(z^\prime)|^2  \right] \Psi(z) \nonumber \\
&+& \frac{1}{2}\left(\lambda - \mathrm{E}_\mathrm{app}\right) \mathrm{d}(z) \psi(\xx,t) + \frac{1}{2}\left( \lambda - \mathrm{E}_\mathrm{app} \right) \sigma(\xx,t) \Psi(z),
\end{eqnarray}
and the equation for $\sigma(\xx,t)$, the linear response of the dipole field, is given by
\begin{eqnarray}
\label{dd}
\sigma(\xx,t) &=& -\alpha \int d\xx^\prime\frac{1-3\cos^2{\theta_{z}}}{|\xx-\xx^\prime|^3}  \left[\mathrm{d}(z^\prime)  \left[ \Psi^\star(z^\prime)\psi(\xx^\prime,t) \right. \right. \nonumber \\
&+& \left. \left. \psi^\star(\xx^\prime,t)\Psi(z^\prime) \right] + \sigma(\xx^\prime,t) |\Psi(z^\prime)|^2\right].
\end{eqnarray}
Now, a complete description of the small fluctuations of the Bose condensate of polar molecules in the linearly polarizable regime must self-consistently satisfy the coupled Eqs.~(\ref{dpsieq}) and~(\ref{dd}).  

In the q2D geometry, the fluctuations $\psi(\xx,t)$ take the form of plane-wave Bogoliubov quasiparticles~\cite{PitaevskiiStringari03},
\begin{equation}
\label{dpsi}
\psi(\xx,t) = \sqrt{n_\mathrm{2D}}\chi(z) \left[ u e^{i\qq \cdot \rrho}e^{-i \omega t} + v^\star e^{-i\qq \cdot \rrho} e^{i\omega t}\right],
\end{equation}
where the $u$ and $v$ amplitudes are normalized to $|u|^2-|v|^2=1$ and we assume that the lowest-lying quasiparticle excitations occupy the axial condensate wave function $\chi(z)$, which amounts to using the single-mode approximation.  Typically, inserting this ansatz into Eq.~(\ref{dpsieq}) results in a coupled set of Bogoliubov de Gennes (BdG) equations for the frequencies $\omega$~\cite{PitaevskiiStringari03}.  Now, these BdG equations have terms $\propto \sigma(\xx,t)$ in addition to the more familiar terms $\propto \psi(\xx,t)$.  Using the BdG ansatz~(\ref{dpsi}), the function $\sigma(\xx,t)$ spatially decouples into radial plane waves and a function that depends only on $z$, much like the form of $\psi(\xx,t)$ in Eq.~(\ref{dpsi}).  Thus, a solution to Eq.~(\ref{dd}) describes a polarization mode with plane-wave character corresponding to a quasiparticle with in-plane momentum $\hbar \qq$.

We solve the modified BdG equations by discretizing them on a numeric grid and using the iterative Arnoldi diagonalization method, solving for $\sigma(\xx,t)$ in Eq.~(\ref{dd}) at each iteration via Gaussian elimination. We take $\chi(z)$ to be a Gaussian normalized to unity with width $l_z$, and find no qualitative and little quantitative difference between this Gaussian ansatz and solving for $\chi(z)$ exactly in the NLSE (Eq.~(\ref{gpe1})).  The solutions are characterized by the quasiparticle dispersion relation, relating the energy $\hbar \omega$ of a quasiparticle and its corresponding polarization mode to its momentum $\hbar \qq$.  As is clear from the ansatz~(\ref{dpsi}), any $\omega$ with a nonzero imaginary part signifies a dynamic instability of the in-plane homogeneous ground state.

To execute the Gaussian elimination algorithm, we represent the integral in Eq.~(\ref{dd}) as a matrix operator in a basis of grid points.  This procedure is stable for the momenta that are relevant for characterizing the stability/phase diagram of the system.  However, for a given $n_\mathrm{2D}\alpha |\chi(0)|^2$, we find larger wave number(s) for which this matrix is singular and the Gaussian elimination fails.  This singularity corresponds to an unphysical divergence in $\sigma(\xx,t)$ and is beyond the scope of our model, in which the dipole moments are limited by a physical cutoff $d_\mathrm{max}$.  The renormalization of this singularity is under active investigation~\cite{RittenhouseUnpub}.  It does not, however, affect our present conclusions.

\begin{table}[h]
\centering
\caption{Candidate Molecules}
\begin{tabular}{c|cccc}
\\[0pt]
Molecule & $\mathrm{d}_\mathrm{max}$ (a.u.) & $\Delta$ (a.u.) & $\alpha$ (a.u.)  & Ref. \\[3pt]
\hline
\\[-5pt]
$^{87}$Rb$^{133}$Cs & 0.49 & $7.6\times 10^{-8}$ & $2.1\times 10^6$ & ~\cite{Kotochigova05,Fellows99} \\[3pt]
$^{232}$Th$^{16}$O & 1.53 & $4.6\times 10^{-12}$ & $1.0\times 10^{12}$ & ~\cite{Buchachenko10,Marian88}  \\[3pt]
\end{tabular}
\label{Candidatemolecules}
\end{table}

To proceed in characterizing this stability/phase diagram, we consider two bosonic molecular species, RbCs and ThO.  The relevant microscopic parameters for these molecules are given in table~\ref{Candidatemolecules}.  While RbCs has a relatively small rotational splitting in its ground state ($\Delta=7.6\times 10^{-8}$ a.u.), ThO has a very small $\Lambda$-doublet splitting in its metastable $H^{3}\Delta_1$ state ($\Delta=4.6\times 10^{-12}$ a.u.).  This results in a very large polarizability for the ThO molecule, and thus enhanced dielectric effects.  These molecules remain in the linearly polarizable regime for applied fields $\mathrm{E}_\mathrm{app} \lesssim \mathrm{d}_\mathrm{max} / (2 \alpha)$, corresponding a maximum applied field of about $200\, \mathrm{V/cm}$ for RbCs and $12\,\mathrm{mV/cm}$ for ThO.  We note that both of these species have been produced in experiments~\cite{Sage05,Vutha10,Debatin11}, although not yet as quantum degenerate gases.

\begin{figure}
\centering
\includegraphics[width=.483\textwidth]{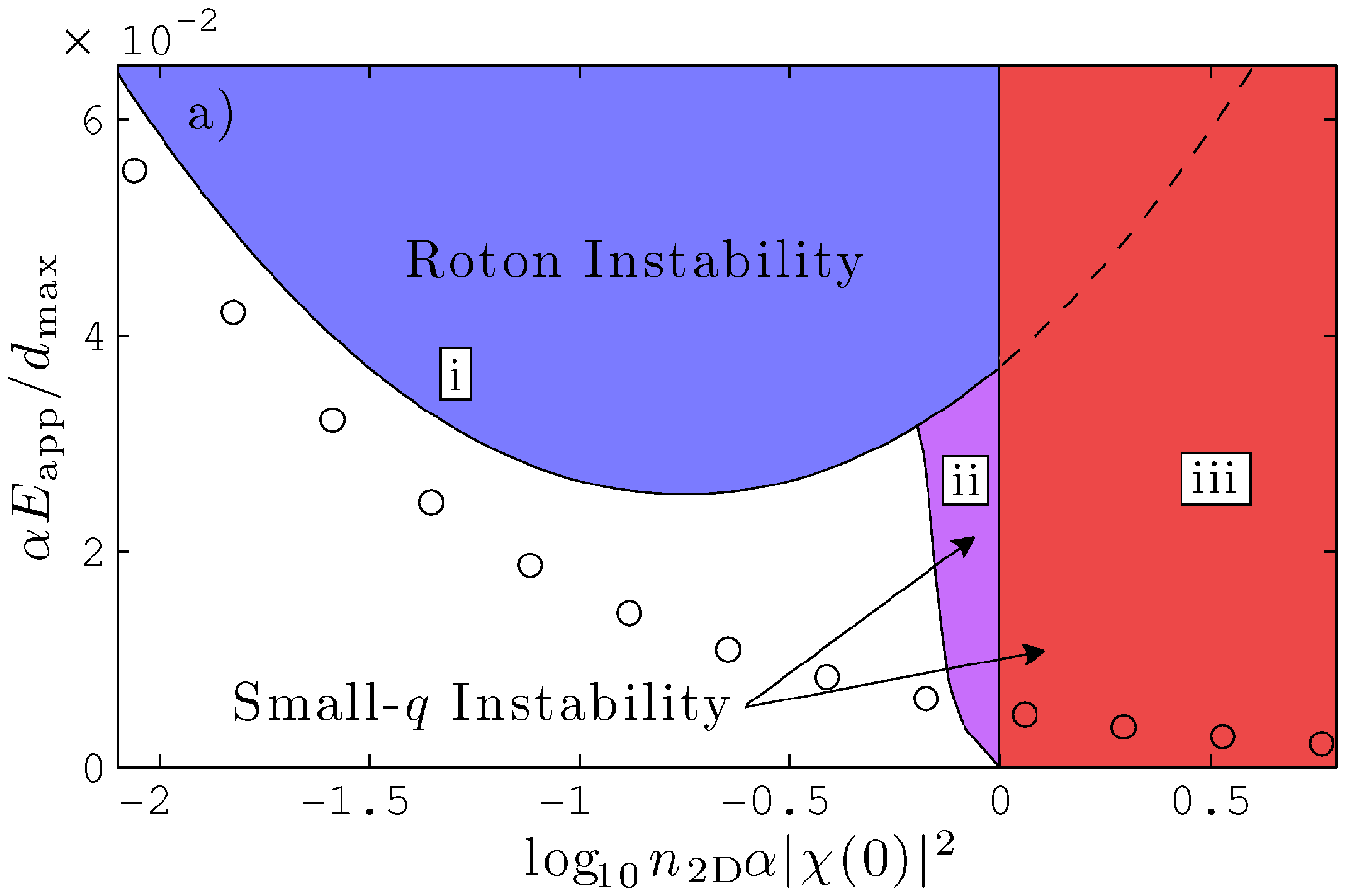}
\includegraphics[width=.497\textwidth]{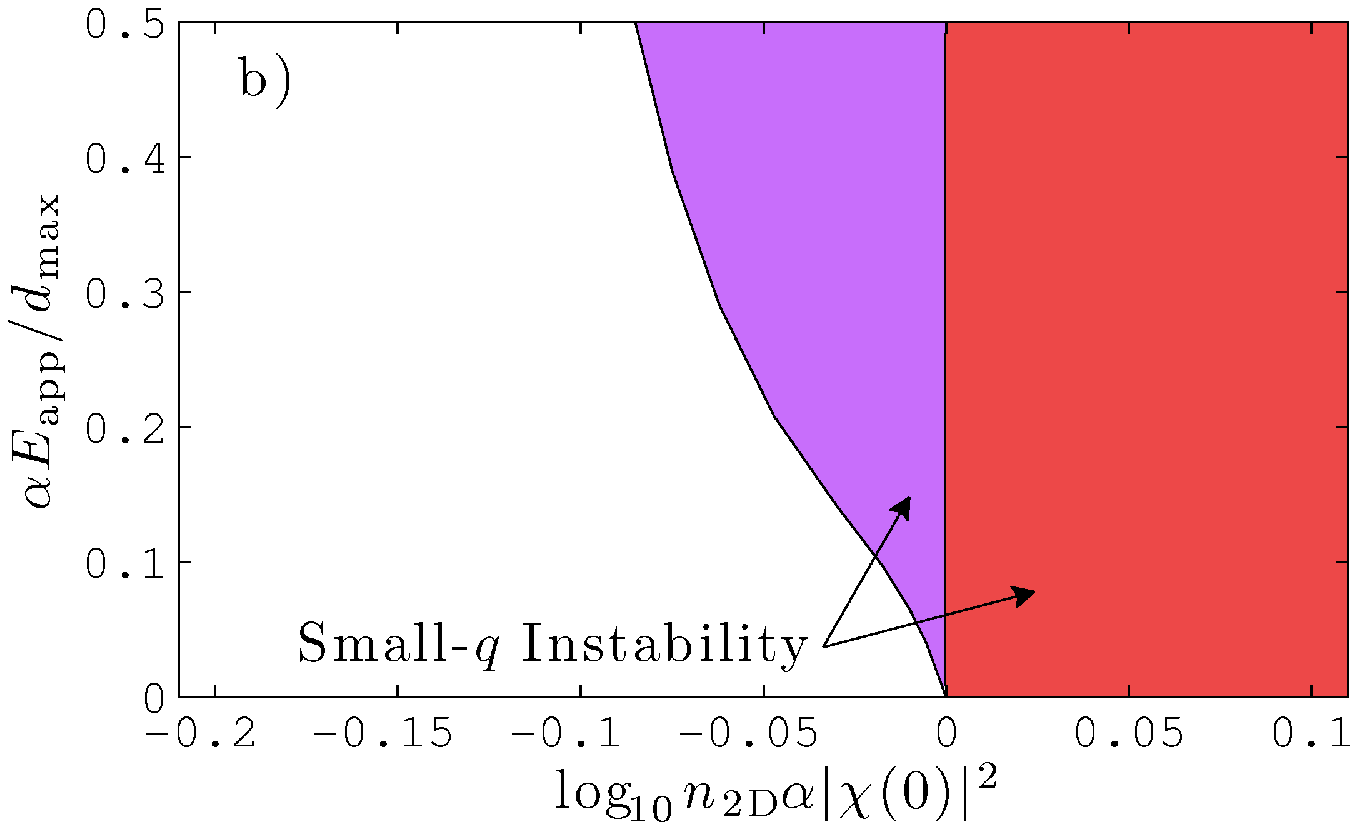} \\
\caption{(color online)  Stability/phase diagram for a) the RbCs system and b) the ThO system in the q2D geometry as a function of applied field $\mathrm{E}_\mathrm{app}$ and integrated 2D density $n_\mathrm{2D}$.  In the RbCs diagram, the blue region (i) signifies roton instability and the purple and red (ii \& iii) shaded regions signify long-wavelength phonon-like instability.  In the red region (iii) there is an instability at all small but nonzero momenta, while in the purple region there is an instability at small momenta, but a gap of stability near zero momentum (a long-wavelength instability with roton character).  The red region (iii) persists for all $\log_{10} n_\mathrm{2D}\alpha |\chi(0)|^2>0$.  The black dashed lines are interpolations of the stability/phase boundaries, and the black circles indicate the applied field above which the q2D dipolar BEC has a roton instability when the effect of $\mathbf{P}(\xx)$ is ignored, i.e., when $ \dd(\xx)=\alpha \EE_\mathrm{app}$.  In the ThO system, the roton instability is completely suppressed and only the phonon-like instabilities are present in the range of applied fields that are permitted in the linearly polarizable regime.}
\label{fig:stab}
\end{figure}

Figures~\ref{fig:stab}(a) and~\ref{fig:stab}(b) characterize the stability/phase of the RbCs system and the ThO system, respectively, as a function of applied field and integrated density.  In~\ref{fig:stab}(a), the blue region (i) shows where the RbCs system has a roton instability due to excitations of finite wavelength $\sim 2\pi l_z$~\cite{Santos03}.  For dipoles of fixed dipole moment, the instability would occur at smaller dipole moment for increasing density, i.e., the boundary of the blue region would be a strictly decreasing function of $n_\mathrm{2D} \alpha$ (shown by the black circles).  However, in the dielectric BEC this trend reverses beyond a critical point and the rotons become more stable, owing to the diminishing dipole moment in the center of the gas (as anticipated in Eq.~(\ref{dz})).  Whereas the roton instability is suppressed in the stability/phase diagram for the RbCs system, it is completely absent from the linearly polarizable ThO system because the large polarizabilities of this molecule serve to suppress the dipole moment in the center of the condensate more rapidly as a function of applied field.  

\begin{figure}
\centering
\includegraphics[width=.8\textwidth]{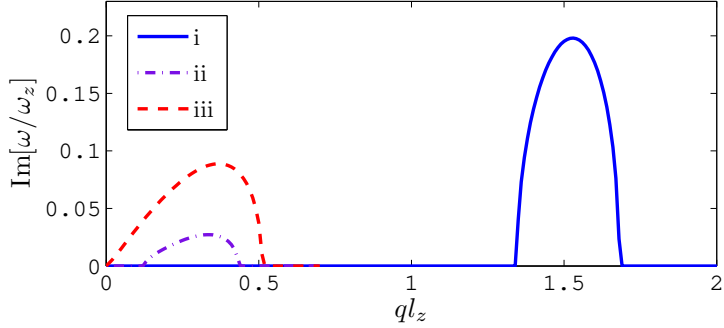} \\
\caption{(color online) The imaginary parts of the dispersions for the parameters that are labeled in figure~\ref{fig:stab}.  The roton instability (i) occurs at larger, finite momentum while the phonon-like dielectric instabilities  (ii \& iii) occur at small momentum.}
\label{fig:disp}
\end{figure}

By contrast, the purple and red (ii \& iii) regions, present in both diagrams~\ref{fig:stab}(a) and~\ref{fig:stab}(b), signify long-wavelength instabilities.  In the purple region (ii), the long-wavelength instability occurs at small but finite momenta, leaving an interval of dynamically stable modes near $q=0$.  As $n_\mathrm{2D}\alpha |\chi(0)|^2$ becomes greater than unity (red region), the instability occurs at \emph{all} small, non-zero momenta, and looks like a phonon instability.  The imaginary parts of these dispersions are given by purple dot-dashed (ii) and the red dashed (iii) lines in figure~\ref{fig:disp}, respectively.  This behavior, being independent of $\mathrm{E}_\mathrm{app}$ and occurring at a critical value of $n\alpha$, is reminiscent of the long-wavelength instability of a homogeneous dielectric with negative static dielectric constant.  {Indeed, this instability signifies the transition to a negative static dielectric function in the q2D geometry, wherein novel physical behavior, such as the attraction of like electric charges, is expected to occur~\cite{Dolgov81}.}

We find that the emergence of the dielectric instability is universal, and occurs at $n_\mathrm{2D}\alpha |\chi(0)|^2=1$ for any linearly polarizable molecule at zero temperature.  For RbCs in a trap with axial frequency $\omega_z = 2\pi\times 20$kHz, this corresponds to a large critical integrated density of $n_\mathrm{2D}=4.02\times 10^{12}$cm$^{-2}$, while for ThO, the critical density is just $7.85\times 10^6$cm$^{-2}$ because these molecules have much larger polarizabilities.  Indeed, the stability/phase diagram of the ThO system in figure~\ref{fig:stab}(b) is identical to that of RbCs in figure~\ref{fig:stab}(a), but with a much smaller cutoff in $\mathrm{E}_\mathrm{app}$ (necessary to remain in the linearly polarizable regime).  Thus, the onset of the dielectric instability occurs at $n_\mathrm{2D}\alpha |\chi(0)|^2=1$ for both species, but the region of roton instability is far suppressed from the ThO diagram due to this molecule's large polarizability.

For the mean-field theory presented in this work to hold, we require that the gas be sufficiently dilute.  In the process of mapping these stability/phase diagrams shown in figure~\ref{fig:stab}, we check that the diluteness criteria is satisfied by calculating the ratio of the characteristic interaction length to the average interparticle spacing across the density profile of the condensate, which is characterized by the 2D gas parameter $n_\mathrm{2D} a_{dd}^2(z)$, where $a_{dd}(z) = M \mathrm{d}^2(z)/3\hbar^2$ is the $z$-dependent characteristic dipole length of a molecule in the system~\cite{Santos03}, varying with the condensate density profile.  We find that this gas parameter is sufficiently small ($\lesssim 10^{-2}$) near the stability/phase boundaries in figure~\ref{fig:stab} to justify our use of a condensate mean-field, while it can become larger ($>1$) for larger densities and applied fields.  In this case, two-body correlations may become important in characterizing the ground state of the gas, however, the dielectric properties of the system should share the qualitative features of the mean field ground state.  Similarly, we expect the qualitative dielectric properties to persist if the molecules possess large Van der Waals lengths~\cite{Kotochigova10}, which can approach the mean intermolecular spacing at sufficiently high density.  

In a realistic experimental setup, a radial harmonic trap is present in addition to the tight axial trap.  Here, long-wavelength quasiparticles manifest as breathing or quadrupole modes.  For the ThO condensate, we consider occupations of $N=500,1000,2000$ molecules in a trap with radial frequency $\omega_\rho=2\pi\times 200$ Hz and the same axial frequency as the q2D system.  We employ the local-density approximation (LDA) to solve the NLSE and calculate the breathing and quadrupole mode frequencies.  This approximation works well to describe these long-wavelength modes in the dipolar system.  We find that, at critical applied fields similar to those in the q2D regime, the breathing and quadrupole modes develop imaginary frequencies, showing that the phonon-like dielectric instability persists in the trapped system, that is, the decay proceeds via the longest wavelength modes available, with a decay time on the order of milliseconds.

To summarize, we have considered a BEC of polar molecules in an applied electric field that is sufficiently weak to keep the dipoles in the linearly polarizable regime.  In this regime, the system develops dielectric character, resulting in a suppressed roton instability and the emergence of a dielectric instability due to the development of a negative static dielectric function.  While this physics is essential to consider for any weak-field studies of molecular BECs, it also introduces new physics that is accessible due to the possibly very large polarizabilities of heteronuclear molecules.  {The molecular BEC gives us access to the regime of the negative static dielectric function, wherein the physics of a superfluid with such novel dielectric properties can be explored for the first time.}

We acknowledge fruitful discussions with Hossein R. Sadeghpour, Dominic Meiser and John Corson.  STR acknowledges the financial support of an NSF grant through ITAMP at Harvard University and the Smithsonian Astrophysical Observatory.  RMW and JLB acknowledge the financial support of the NSF.

\end{document}